\documentclass[aps,twocolumn,preprintnumbers,amsmath,amssymb,nofootinbib,superscriptaddress,notitlepage]{revtex4}

\usepackage{epsfig}
\usepackage{slashed}

\usepackage{multirow}

\begin{document}

\title{The $D^* D^* \bar{D}$ and $D^* D^* \bar{D}^*$ Three-Body Systems}

	\author{M. Pavon Valderrama}\email{mpavon@buaa.edu.cn}
\affiliation{School of Physics and Nuclear Energy Engineering, \\
International Research Center for Nuclei and Particles in the Cosmos and \\
Beijing Key Laboratory of Advanced Nuclear Materials and Physics, \\
Beihang University, Beijing 100191, China} 

\date{\today}


\begin{abstract} 
  \rule{0ex}{3ex}
  The hidden charm $X(3872)$ resonance is usually thought
  to be a $D^{*0} \bar{D}^0$ meson-antimeson molecule
  with quantum numbers $J^{PC} = 1^{++}$.
  If this is the case, there is the possibility that there might be
  three body bound states with two charmed mesons
  and a charmed antimeson.
  Here we argue that the theoretical existence of this type of
  three body molecules is expected from heavy quark spin symmetry.
  If applied to the two body sector, this symmetry implies that the interaction
  of the $D^{*0} \bar{D}^{*0}$ meson-antimeson pair
  in the $J^{PC} = 2^{++}$ channel
  is the same as in the $J^{PC} = 1^{++}$ $D^{*0} \bar{D}^0$ case.
  From this we can infer that the $J^P = 3^{-}$ $D^{*0} D^{*0} \bar{D}^{*0}$
  molecule will be able to display the Efimov effect if the scattering length of
  the $2^{++}$ channel is close enough to the unitary limit.
  Heavy quark spin symmetry also indicates that
  the $J^P = 2^{-}$ $D^{*0} D^{*0} \bar{D}^0$ molecule is analogous to
  the $J^P = 3^{-}$ $D^{*0} D^{*0} \bar{D}^{*0}$ one.
  That is, it can also have a geometric spectrum.
  If we consider these triply heavy trimers in the isospin symmetric limit,
  the Efimov effect disappears and we can in principle predict
  the fundamental state of the $2^-$ $D^*D^*\bar{D}$ and
  $3^-$ $D^* D^* \bar{D}^*$ systems.
  The same applies to the $B^*B^* \bar{B}^*$ system:
  if the $Z_b(10650)$ is an isovector $B^* \bar{B}^*$ molecule then
  the $0^-$ isodoublet and the $1^-$, $2^-$ isoquartet $B^* B^* \bar{B}^*$
  trimers might bind, but do not display Efimov physics.
  Finally from heavy flavour symmetry it can be argued that
  scattering in the $B D$ two-body system might be resonant.
  This would in turn imply the possibility of Efimov physics
  in the $B B D$ three body system.
\end{abstract}

\maketitle

\section{Introduction}

The three boson system in the unitary limit, i.e. when the two-body scattering
length goes to infinity, shows a geometric spectrum in which the ratio
of the energy of the $n$-th and $(n+1)$-th excited trimer is 
$E_{n} / E_{n+1} \simeq 521$.
This spectrum, originally theorized by Efimov
in the seventies~\cite{Efimov:1970zz},
has been confirmed experimentally
with cesium atoms a decade ago~\cite{Kraemer:2006}.
The relevance of the Efimov effect~\cite{Hammer:2010kp,Naidon:2016dpf}
is not limited to atomic physics,
but extends to nuclear physics where it might play
an important role in the binding of the triton~\cite{Bedaque:1998kg,Bedaque:1998km,Bedaque:1999ve},
the three-alpha structure of Carbon-12~\cite{Hammer:2008ra} or
two-neutron halo nuclei~\cite{Federov:1994cf,Hammer:2017tjm}.
The Efimov effect also extends to other three-body systems
in the unitary limit:
relevant to the present investigation is the case of two identical
non-interacting boson of species $A$ and a third particle of
species $B$ that interacts resonantly with the other two,
i.e. the $AB$ scattering length diverges.
This system displays a geometric spectrum too~\cite{Efimov:1970zz}.
If $m_A$ and $m_B$ are the masses of particles of species $A$ and $B$
respectively, the larger the mass imbalance $m_A / m_B$
the more conspicuous will it be the geometric spectrum.
For $m_A = m_B$ the scaling factor between two consecutive states is
$E_{n} / E_{n+1} \simeq (1986.1)^2$~\cite{Helfrich:2010yr}.
The existence of a geometric spectrum can also happen in specific
two-body systems that fulfill a series of properties~\cite{Geng:2017hxc}.
A remarkable example is the $\bar{D}^* \Sigma_c$-$\bar{D} \Lambda_c$
system, which might be identified with the $P_c(4450)$ pentaquark-like
resonance~\cite{Aaij:2015tga} in the molecular picture~\cite{Chen:2015loa,Chen:2015moa,Roca:2015dva,He:2015cea,Xiao:2015fia}.

There is the interesting question of whether there are three hadron systems
(besides the three nucleon system) where the Efimov effect
might play a role~\cite{Canham:2009zq,Wilbring:2017fwy,Raha:2017ahu,Valderrama:2018knt}.
This is particularly relevant in view of the recent renaissance of hadron
spectroscopy triggered by the discovery of the $X(3872)$
by the Belle collaboration more than one decade ago~\cite{Choi:2003ue}.
The $X(3872)$ is a $J^{PC} = 1^{++}$ narrow resonance that is located within
current experimental uncertainties on top of the $D^0 D^{0*}$ threshold.
This striking coincidence naturally leads to the interpretation that
the $X(3872)$ is a $D^0 D^{0*}$ bound state~\cite{Tornqvist:2003na,Voloshin:2003nt,Braaten:2003he}.
The strongest hint that the $X(3872)$ is molecular lies in the branching
ratio of the isospin breaking decays $\Gamma(X \to J/\Psi 2 \, \pi)$
and $\Gamma(X \to J/\Psi \, 3 \pi)$~\cite{Choi:2011fc},
which are easy to explain
in the molecular picture~\cite{Gamermann:2009fv,Gamermann:2009uq}
but are more problematic if the $X(3872)$
is a more compact object~\cite{Hanhart:2011tn}.
This does not preclude the existence of a small $c\bar{c}$ component,
negligible at long distances but important for explaining
the radiative charmonia decays $X \to J/\Psi \gamma$ and
$X \to \Psi(2S) \gamma$~\cite{Aaij:2014ala}
which probe the structure of the $X(3872)$
at short distances~\cite{Swanson:2004pp,Guo:2014taa}.

If the $X(3872)$ is indeed a bound state its binding energy will be extremely
small and the scattering length extremely large~\cite{Braaten:2003he}.
That is, a molecular $X(3872)$ will be close to the unitary limit.
This has prompted the question of whether there will be an Efimov-like
geometric spectrum if we add a third charmed meson.
In this regard, Canham, Hammer and Springer~\cite{Canham:2009zq}
considered $D^0 X$ and $D^{*0} X$ scattering from effective field theory
and pointed out that the system does not fulfill
the conditions to have a geometric Efimov-like spectrum.
The reason is the existence of several channels in which the interaction
is not resonant, which dilutes the interaction strength provided by
the $1^{++}$ $D^0 \bar{D}^{*0}$ channel.
In the present manuscript we update the previous conclusions
regarding the $D^{0*} X$ system.
We notice that if we consider heavy quark spin symmetry then
$D^{*0} \bar{D}^{*0}$ scattering in the $J^{PC} = 2^{++}$ channel
might also be resonant.
If this additional channel happens to be resonant, it will provide enough
attraction as to trigger the Efimov effect in the $2^-$ $D^{0*} X$ system and
the $3^-$ $D^{0*} D^{0*} \bar{D}^{0*}$ three body system.
Conversely if we turn off the $2^{++}$ $D^{*0} \bar{D}^{*0}$ scattering channel
we recover the conclusions of Ref.~\cite{Canham:2009zq}.

In the isospin symmetric limit the $X(3872)$ can be considered
to be an isoscalar $1^{++}$ $D^*\bar{D}$ molecule with a binding energy
of about $4\,{\rm MeV}$.
In this limit there is no Efimov effect,
independently of the location of the bound state.
The reason is the loss of attraction owing to the numerical factors
involved in the coupling of isospin.
Only if the $1^{++}$ $D^* \bar{D}$ and $2^{++}$ $D^* \bar{D}^*$ interactions
were resonant in the isoscalar and isovector channels
will the Efimov effect be present.
The absence of Efimov physics makes it possible to make predictions
for the $2^-$ $D^* D^* \bar{D}$ and $3^{-}$ $D^* D^* \bar{D}^*$ systems
without including a three body force.
For these two systems we find a three-body binding energy of about
$B_3 \simeq 2-3\,{\rm MeV}$, where this number is defined
with respect to the dimer-particle threshold.
If we refer to the three body threshold instead, the $2^-$ and $3^-$ states
will be located $7-9\,{\rm MeV}$ below it.

The $Z_b(10610)$ and $Z_b(10650)$ isovector resonances have been speculated
to be $I^G(J^{PC}) = 1^+(1^{+-})$ $B^* \bar{B}$ and $B^* \bar{B}^*$ molecules.
If this is the case the previous ideas also apply {\it mutatis mutandis}
to the $B^* B^* \bar{B}^*$ system.
The molecules that can be predicted from the $Z_b(10650)$ are a trio of
degenerate trimers: a $0^-$ isodoublet and a $1^-$ and $2^-$ isoquartet.
Their existence is contingent on the location of the $Z_b(10650)$
as a bound state: for a two-body binding energy of $B_2 = 2\,{\rm MeV}$,
the trimer binding will be $B_3 \simeq 1\,{\rm MeV}$.

The Efimov effect is more probable for systems with a mass imbalance,
as we have already mentioned.
This idea can be combined with heavy flavour symmetry, from which we can
theorize the existence of $B^+ D$, $B^{*+} D^0$, $B^+ D^{*0}$ and $B^{*+} D^{*0}$
molecules~\cite{Guo:2013sya}.
If these $\bar{b} c$-like molecules exists and are close to the unitary limit,
the $B^+ B^+ D^0$, $B^+ B^{*+} D^0$, $B^{*+} B^{*+} D^0$ and $B^+ B^+ D^{*0}$
family of three heavy meson system might be a likely candidate
for a geometric spectrum.
The analysis of the Faddeev equations for these family of trimers indicates
that the appearance of the Efimov effect is possible.
The discrete scaling factor for the binding energies turns out to be
$E_{n+1}/E_n \sim 4000-5000$, where the exact number depends on the
particular system under consideration.

The manuscript is structured as follows: after the introduction,
we briefly comment on the heavy-light spin decomposition of
the three heavy meson states in Section \ref{sec:HL}.
After that we will present the Faddeev equations for these systems
in Section \ref{sec:Faddeev}.
We will analyze the conditions for the existence of the Efimov effect
in Section \ref{sec:Efimov}.
We will present the results for the $2^-$ $D^* D^* \bar{D}$ and
$3^{-}$ $D^* D^* \bar{D}^*$ systems in Section \ref{sec:DDD}.
The isodoublet $0^{-}$ and isoquartet $1^-$ and $2^-$ $B^*B^* \bar{B}^*$ trimers
are studied in Section \ref{sec:BBB}.
Then we will extend the formalism to the $BBD$, $BB^*D$, $B^*B^*D$
and $BBD^*$ molecules in Section \ref{sec:BBD}.
Finally we will present the conclusions of this work at the end.

\section{Heavy-Light Spin Decomposition of the $HH\bar{H}$ System}
\label{sec:HL}

From heavy quark spin symmetry (HQSS) we expect the spectrum of hadron
systems containing a mixture of heavy ($Q = c,b$) and light ($q = u,d,s$)
degrees of freedom to be independent of the spin of
the heavy quarks~\cite{AlFiky:2005jd,Bondar:2011ev,Mehen:2011yh,Valderrama:2012jv,Nieves:2012tt}.
Conversely, the spectrum of these systems mostly depends
on the total angular momentum of the light quarks.
We can state this idea in more concrete terms by considering the heavy-light
decomposition of a system of $N$ heavy hadrons $H_1$, $H_2$, $\dots$, $H_N$
with total angular momentum $J$
\begin{eqnarray}
  | H_1 H_2 \dots H_N (J)\rangle =
  \sum_{J_H, J_L} c(J_H, J_L)\,J_H \otimes J_L \Big|_J \, ,
\end{eqnarray}
where the $c(J_H, J_L)$ are the coefficients involved
in this angular momentum decomposition~\footnote{Actually the decomposition
  is more involved than stated here: each $J_H$ and $J_L$ can be further
  decomposed in orthogonal contributions corresponding to different
  intermediate couplings of the angular momenta. For example, if we have
  three heavy hadrons $J_H = \frac{1}{2}$ can be further decomposed
  into $0_{12} \otimes \frac{1}{2}$ and $1_{12} \otimes \frac{1}{2}$,
  where $0_{12}$ and $1_{12}$ refer to coupling of
  heavy hadrons $1$ and $2$.
  Yet for the current illustrative purposes
  this distinction will not be made.}.
For this system {\it loosely speaking} the energy levels can be determined
in terms of the total light quark angular momenta
\begin{eqnarray}
  E_N = \sum d(J_L)\,E(J_L) \, ,
\end{eqnarray}
with
\begin{eqnarray}
  d(J_L) = \sum_{J_H}\,|c(J_H, J_L)|^2 \, .
\end{eqnarray}
The reason why we say {\it loosely speaking} is because for a non-relativistic
system of $N$ heavy hadrons what is determined from this decomposition
is the potential energy rather than the binding energy,
yet the idea can be illuminating.

If we consider for instance the charmed meson-antimeson system,
the spectrum is determined by the fact that
the total light spin is either $J_L = 0$ or $J_L = 1$.
For instance, the $D^* \bar{D}^*$ system with $J^{PC} = 2^{++}$ can be trivially
decomposed as
\begin{eqnarray}
  | D^* \bar{D}^* (2^{++}) \rangle = 1_H \otimes 1_L \Big|_{J=2} \, .
\end{eqnarray}
For the $D \bar{D}^*$/$D^* \bar{D}$ system we have exactly
the same decomposition for the $J^{PC} = 1^{++}$ state~\footnote{Notice
  that here we are implicitly taking the C-parity convention
  $C | D^* \rangle = | \bar{D}^* \rangle$. This will lead to a far simpler
  analysis in the three body sector.}
\begin{eqnarray}
  | D \bar{D}^* (1^{++}) \rangle &=&
  \frac{1}{\sqrt{2}}\,| D \bar{D}^* (1^{+}) \rangle +
  \frac{1}{\sqrt{2}}\,| D^* \bar{D} (1^{+}) \rangle \, , \nonumber \\
  &=& 1_H \otimes 1_L \Big|_{J=1} \, ,
\end{eqnarray}
which corresponds to the $X(3872)$.
From this we deduce that the the interaction in the $J_L = 1$ channel
is attractive and leads to the formation of a bound state.
We also deduce that if the $1^{++}$ system binds, the same should be true
for the $2^{++}$ system as the interaction is identical.
Of course this conclusion is subjected to the limitation
that HQSS is violated at the $\Lambda_{QCD} / m_c \sim 10-15\,\%$ level
in the charm sector plus the fact that the $X(3872)$ is barely bound.
If we take this into account we can still expect the $2^{++}$ interaction
to be strong, but not necessarily resonant or leading to a bound state.

This idea can be trivially extended to the three charmed meson system,
in which case the total light angular momentum is either $J_L = \frac{1}{2}$
or $J_L = \frac{3}{2}$.
For the quantum numbers $J^{P} = 3^{-}$ the decomposition is indeed trivial
\begin{eqnarray}
  | D^{*} D^{*} \bar{D}^{*} (3^-) \rangle &=&
  {\frac{3}{2}}_H \otimes {\frac{3}{2}}_L \Big|_{J=3} \, .
\end{eqnarray}
There is also a $J^{P} = 2^{-}$ system for which the decomposition is identical
\begin{eqnarray}
  \phantom{+}
  \frac{1}{\sqrt{3}}\,| D^{*} D^{*} \bar{D} (2^-) \rangle && \nonumber \\
  +\frac{1}{\sqrt{3}}\,| D^{*} D \bar{D}^{*} (2^-) \rangle && \nonumber \\
  +\frac{1}{\sqrt{3}}\,| D D^{*} \bar{D}^{*} (2^-) \rangle &=&
  {\frac{3}{2}}_H \otimes {\frac{3}{2}}_L \Big|_{J=2} \, , 
\end{eqnarray}
and this implies that if the $3^{-}$ trimer binds,
the $2^{-}$ trimer should also bind.

\section{Faddeev Equations for the $HH\bar{H}$ System }
\label{sec:Faddeev}

In this section we write the Faddeev decomposition and equations for
the $3^{-}$ $D^* D^* \bar{D}^*$, $2^{-}$ $D^* D^* \bar{D}^*$ and
$1^{-}$ $DD\bar{D}^*$ charmed meson-meson-antimeson trimers.
We will consistently assume that the $D D$, $D D^*$ and $D^* D^*$ 
charmed meson-meson pairs do not interact.
Equivalently, we consider that their interaction is weak~\footnote{
If we adapt the OBE model of Ref.~\cite{Liu:2018bkx} to the case at hand,
the isovector $0^+$ $D D$, $1^+$ $D D^*$ and $2^+$ $D^* D^*$ scattering lengths
happen to be $a_0 = -0.1\,{\rm fm}$, $-0.6\,{\rm fm}$ and $-0.7\,{\rm fm}$.
That is, these channels are slightly attractive and might increase
the trimer binding energy by a very small amount.
}
and will only provide a small subleading correction
to the three body binding energy.

\subsection{The Equations for $D^* D^* \bar{D}^*$}

We begin by writing the three body wave function in terms of Faddeev components
for the $J^P=3^{-}$ $D^{*0} D^{*0} \bar{D}^{*0}$ system
\begin{eqnarray}
  \Psi_{3B} = \left[ \phi(\vec{k}_{23},\vec{p}_1) + \phi(\vec{k}_{31},\vec{p}_2)
    \right] \, | D^{*0} D^{*0} \bar{D}^{*0} \rangle \, , \nonumber \\
\end{eqnarray}
where we assume that the $D^{0*} D^{0*}$ subsystem is not interacting,
from which we can ignore the third component of the Faddeev decomposition.
The Jacobi momenta are defined as usual
\begin{eqnarray}
  \vec{k}_{ij} &=& \frac{m_j \vec{k}_i - m_i \vec{k}_j}{m_i + m_j} \, , \\
  \vec{p}_{k} &=& \frac{1}{M_T}\,\left[ (m_i + m_j)\,\vec{k}_k -
    m_k\,(\vec{k}_i + \vec{k}_j) \right] \, , 
\end{eqnarray}
with $m_1$, $m_2$, $m_3$ the masses of particles $1$, $2$, $3$,
$M_T = m_1 + m_2 + m_3$ the total mass and $ijk$ an even permutation of $123$.
In this case we have $m_1 = m_2 = m_3 = m_{D^{*0}}$. 
If the charmed mesons interact via a potential of the type
\begin{eqnarray}
  V_{D^* \bar{D}^*}(J=2) = C_0 g(p) g(p') \, ,  
\end{eqnarray}
where $g(p)$ is a regulator function then
the $D^* \bar{D}^*$ T-matrix can be written as
\begin{eqnarray}
  T_{23} = \tau(Z) g(p) g(p') \, ,
\end{eqnarray}
while the Faddeev component $\phi$ admits the ansatz
\begin{eqnarray}
  \phi(k,p) = \frac{g(k)}{Z - \frac{k^2}{2 \mu_{23}} - \frac{p^2}{2 \mu_1}} a(p)
  \, ,
\end{eqnarray}
with the reduced masses defined as
\begin{eqnarray}
  \frac{1}{\mu_{ij}} &=& \frac{1}{m_i} + \frac{1}{m_j} \, , \\
  \frac{1}{\mu_{k}} &=& \frac{1}{m_k} + \frac{1}{m_i + m_j} \, .
\end{eqnarray}
From the previous, we find that $a(p)$ follows the integral equation
\begin{eqnarray}
  a(p_1) = \tau_{2}(Z_{23})\,\int \frac{d^3 p_2}{(2\pi)^3}\,
  B^0_{12}(\vec{p}_1, \vec{p}_2)\,a(p_2) \, ,
\end{eqnarray}
where the driving term $B^0_{12}$ is given by
\begin{eqnarray}
  B^0_{ij} (\vec{p}_i, \vec{p}_j) =
  \frac{g(q_i)\,g(q_j)}
       {Z - \frac{p_1^2}{2 m_1} - \frac{p_2^2}{2 m_2} - \frac{p_3^2}{2 m_3}} \, ,
\end{eqnarray}
with $\vec{p}_1 + \vec{p}_2 + \vec{p}_3 = 0$ and
\begin{eqnarray}
  \vec{q}_k = \frac{m_j \vec{p}_i - m_i \vec{p}_j}{m_j + m_i} \, .
\end{eqnarray}
The integral equation can be discretized, in which case finding
the bound state solution reduces to an eigenvalue problem.

\subsection{The Equations for $D^* D^* \bar{D}$}

The Faddeev decomposition of the three body wave function
for the $J^P=2^-$ $D^* D^* \bar{D}$ system involves
three different particle channels
\begin{eqnarray}
  \Psi_{3B} &=&
  \left[ \phi_1(\vec{k}_{23},\vec{p}_1) + \phi_1(\vec{k}_{31},\vec{p}_2)
    \right] \, | D^{*0} D^{*0} \bar{D}^{0} \rangle \nonumber \\
  &+& \left[ \phi_2(\vec{k}_{23},\vec{p}_1) + \varphi_2(\vec{k}_{31},\vec{p}_2)
    \right] \, | D^{*0} D^{0} \bar{D}^{*0} \rangle \nonumber \\
  &+& \left[ \varphi_2(\vec{k}_{23},\vec{p}_1) + \phi_2(\vec{k}_{31},\vec{p}_2)
    \right] \, | D^{0} D^{*0} \bar{D}^{*0} \rangle \, , \nonumber \\
\end{eqnarray}
but it is essentially analogous to that of the $J^P=2^{-}$ $D^* D^* \bar{D}^*$
system.
Notice that we do not explicitly write the spin wave functions.
The choice of wave functions is constrained by the symmetry of the first
particle channel, i.e. $| D^{*0} D^{*0} \bar{D}^{0} \rangle$.
Particles $1$ and $2$ are considered to be non-interacting,
while the interaction with particle $3$ (the charmed anti-meson)
can be written as
\begin{eqnarray}
  \langle D \bar{D}^* | T_{23}(Z) | D \bar{D}^* \rangle &=&
  \tau_{1D}(Z) g(p) g(p')
  \, , \\
  \langle D \bar{D}^* | T_{23}(Z) | D^* \bar{D} \rangle &=&
  \tau_{1E}(Z) g(p) g(p')
  \, ,
\end{eqnarray}
for the $D\bar{D}^*$ system, while for the $D^*\bar{D}^*$ case we write
\begin{eqnarray}
  \langle D^* \bar{D}^* | T_{23}(Z) | D^* \bar{D}^* \rangle &=&
  \tau_{2}(Z) g(p) g(p')
  \, .
\end{eqnarray}
That is, for the $D\bar{D}^*$ we are distinguishing between a direct scattering
process, where the $D$ and $D^*$ are not exchanged,
and an exchange scattering process, where the $D$ and $D^*$ are flipped.

The three independent Faddeev components $\phi_1$, $\phi_2$, and $\varphi_2$
can be written as
\begin{eqnarray}
  \phi_1(k,p) &=&
  \frac{g(k)}{Z - \frac{k^2}{2 \mu_{23}} - \frac{p^2}{2 \mu_1}} a_1(p)
  \, , \\
  \phi_2(k,p) &=&
  \frac{g(k)}{Z - \frac{k^2}{2 \mu_{23}} - \frac{p^2}{2 \mu_1}} a_2(p)
  \, , \\
  \varphi_2(k,p) &=&
  \frac{g(k)}{Z - \frac{k^2}{2 \mu_{31}} - \frac{p^2}{2 \mu_2}} b_2(p)
  \, ,
\end{eqnarray}
where for $\phi_1$ the ordering of particles is $| 123 \rangle = | D^{*0} D^{*0} \bar{D}^{0} \rangle $, while for $\phi_2$ and $\varphi_2$ is
$| 123 \rangle = | D^{*0} D^{0} \bar{D}^{*0} \rangle$.
The corresponding Faddeev equations are
\begin{eqnarray}
  a_1(p_1) &=& \tau_{1D}(Z_{23})\,\int \frac{d^3 \vec{p}_2}{(2 \pi)^3}\,
  B^0_{12}(\vec{p}_1, \vec{p}_2)\,a_1(\vec{p}_2) \nonumber \\
  &+& \tau_{1E}(Z_{23})\,\int \frac{d^3 \vec{p}_2}{(2 \pi)^3}\,
  B^0_{12}(\vec{p}_1, \vec{p}_2)\,b_2(\vec{p}_2)  \, , \\
  a_2(p_1) &=& \tau_{1E}(Z_{23})\,\int \frac{d^3 \vec{p}_2}{(2 \pi)^3}\,
  B^0_{12}(\vec{p}_1, \vec{p}_2)\, a_1(p_2) \nonumber \\
  &+& \tau_{1D}(Z_{23})\,\int \frac{d^3 \vec{p}_2}{(2 \pi)^3}\,
  B^0_{12}(\vec{p}_1, \vec{p}_2)\, b_2(p_2)  \, , \\
  b_2(p_1) &=& \tau_{2}(Z_{13})\,\int \frac{d^3 \vec{p}_2}{(2 \pi)^3}\,
  B^0_{12}(\vec{p}_1, \vec{p}_2)\, a_2(p_2) \, .
\end{eqnarray}
These equations can be simplified if we consider a series of properties
of the system.
First, we can write the direct and exchange T-matrices as combinations
of the T-matrices in the $D\bar{D}^*$ positive and negative C-parity
channels
\begin{eqnarray}
  \tau_{1D} &=& \frac{1}{2} \tau_1^{+} + \frac{1}{2} \tau_1^{-}  \, , \\
  \tau_{1E} &=& \frac{1}{2} \tau_1^{+} - \frac{1}{2} \tau_1^{-} \, ,
\end{eqnarray}
where $\tau_1^{\pm}$ refers to the scattering in the $C = \pm 1$ channel.
Second, we will assume that the interaction in the positive C-parity $X(3872)$
channel is strong while the interaction in the negative C-parity channel is
weak and can be ignored.
In this case we end up with
\begin{eqnarray}
  \tau_{1D} &=& \tau_{1E} = \frac{1}{2} \tau_1^{+} \, .
\end{eqnarray}
from which it also follows that
\begin{eqnarray}
  a_1 = a_2 \, . 
\end{eqnarray}
Third, from HQSS we also expect that
\begin{eqnarray}
  \tau_1^{+} = \tau_2 \, ,
\end{eqnarray}
from which we obtain that
\begin{eqnarray}
  b_2 = a_1 \, .
\end{eqnarray}
With all these relations and if we ignore the mass difference between the
$D^0$ and $D^{*0}$ mesons, the Faddeev equations reduce to
\begin{eqnarray}
  a_1(p_1) &=& \tau_2(Z_{23})\,\int \frac{d^3 \vec{p}_2}{(2 \pi)^3}\,
  B^0_{12}(\vec{p}_1, \vec{p}_2)\,a_1(p_2)  \, , 
\end{eqnarray}
that is, exactly the same Faddeev equation that we had for the $J^P=3^-$
$D^{*0} D^{*0} \bar{D}^{*0}$ system.

Instead of the constructive approach that we have followed before,
alternatively we could have simply begun with the HQSS
three body wave function with $J_L = \frac{3}{2}$,
that is
\begin{eqnarray}
  \Psi_{3B} &=&
  \left[ \phi_1(\vec{k}_{23},\vec{p}_1) + \phi_1(\vec{k}_{31},\vec{p}_2)
    \right] \, | D^{*0} D^{*0} \bar{D}^{0} \rangle \nonumber \\
  &+& \left[ \phi_1(\vec{k}_{23},\vec{p}_1) + \phi_1(\vec{k}_{31},\vec{p}_2)
    \right] \, | D^{*0} D^{0} \bar{D}^{*0} \rangle \nonumber \\
  &+& \left[ \phi_1(\vec{k}_{23},\vec{p}_1) + \phi_1(\vec{k}_{31},\vec{p}_2)
    \right] \, | D^{0} D^{*0} \bar{D}^{*0} \rangle \, , 
\end{eqnarray}
in which case we will had obtained the same eigenvalue equation,
only in a more direct manner.

\subsection{The Equations for $D^0 D^0 \bar{D}^{0*}$}

For the $J^P=1^-$ $D^0 D^0 \bar{D}^{0*}$ system
the non-trivial Faddeev components can be written as
\begin{eqnarray}
  \Psi_{3B} &=&
  \left[ \phi_1(\vec{k}_{23},\vec{p}_1) + \phi_1(\vec{k}_{31},\vec{p}_2)
    \right] \, | D^0 D^{0} \bar{D}^{*0} \rangle \nonumber \\
  &+& \phi_2(\vec{k}_{23},\vec{p}_1) 
  \, | D^{0} D^{*0} \bar{D}^{0} \rangle \nonumber \\
  &+& \phi_2(\vec{k}_{31},\vec{p}_2)
  \, | D^{*0} D^{0} \bar{D}^{0} \rangle \, ,
\end{eqnarray}
where we are assuming that the charmed meson-meson and the $D^0 \bar{D}^0$
interactions are both trivial.
The ansatz for the $\phi_1$ and $\phi_2$ wave functions is identical
to the $J^P=2^-$ $D^{*0} D^{*0} \bar{D}^{0}$ system,
from which we derive the equations
\begin{eqnarray}
  a_1(p_1) &=& \tau_{1D}(Z_{23})\,\int \frac{d^3 \vec{p}_2}{(2 \pi)^3}\,
  B^0_{12}(\vec{p}_1, \vec{p}_2)\,a_1(p_2) \, , \\
  a_2(p_1) &=& \tau_{1E}(Z_{23})\,\int \frac{d^3 \vec{p}_2}{(2 \pi)^3}\,
  B^0_{12}(\vec{p}_1, \vec{p}_2)\,a_1(p_2) \, .
\end{eqnarray}
We can see that the equation in the first line is homogeneous
while the one in second line is inhomogeneous,
i.e. $a_1(p)$ can be determined by itself,
while $a_2(p)$ can be determined from $a_1(p)$.
In other words, the binding energy of the charmed meson trimers
can be determined exclusively from the homogeneous equation.

If we now take into account that we only expect the interaction to be resonant
in the positive C-parity channel, we can rewrite the equations as
\begin{eqnarray}
  a_1(p_1) &=& \frac{1}{2}\tau^{+}(Z_{23})\,\int \frac{d^3 \vec{p}_2}{(2 \pi)^3}\,
  B^0_{12}(\vec{p}_1, \vec{p}_2)\, a_1(p_2) \, , \\
  a_2(p_1) &=& \frac{1}{2}\tau^{+}(Z_{23})\,\int \frac{d^3 \vec{p}_2}{(2 \pi)^3}\,
  B^0_{12}(\vec{p}_1, \vec{p}_2)\, a_1(p_2) \, .
\end{eqnarray}
This will be later very useful to understand why we do not expect
the $J^P=1^-$ $D^{0} D^{0} \bar{D}^{*0}$ system
to have a geometric Efimov-like spectrum.

\subsection{The Isospin Symmetric Limit}

Finally we consider the previous three body molecules in the isospin symmetric
limit.
For the $J^P=3^{-}$ $D^{*} D^{*} \bar{D}^{*}$ molecule
the Faddeev decomposition is
\begin{eqnarray}
  \Psi_{3B} = \left[ \phi(\vec{k}_{23},\vec{p}_1) + \phi(\vec{k}_{31},\vec{p}_2)
    \right] \, | D^{*} D^{*} \bar{D}^{*} \rangle
  | 1 \otimes \frac{1}{2} \rangle_{\frac{1}{2}} \, , \nonumber \\
\end{eqnarray}
where in addition to the particle wave function we have included
the isospin wave function
\begin{eqnarray}
  | I_{12} \otimes I_3 \rangle_{I_T} \, ,
\end{eqnarray}
which means that particles $1$ and $2$ couple to isospin $I_{12}$, particle
$3$ has isospin $I_3$ and the total isospin of the system is $I_T$.
The choice $I_{12} = 1$ together with the fact that the total spin is $S=3$
(and hence $S_{12} = 2$) implies that the spin and isospin wave
functions are symmetric.
In the isospin symmetric limit we expect the $1^{++}$ $D \bar{D}^*$ and
$2^{++}$ $D^* \bar{D}^*$ interactions to be resonant
in the isoscalar $I=0$ channel and
weak in the isovector channel.
If we ignore the isovector interaction, the Faddeev equation reduces to
\begin{eqnarray}
  a(p_1) = \frac{3}{4}\tau^{\rm IS}_2(Z_{23})\,\int \frac{d^3 p_2}{(2\pi)^3}\,
  B^0_{12}(\vec{p}_1, \vec{p}_2)\,a(p_2) \, ,
\end{eqnarray}
where the $3/4$ is an isospin factor which is derived from overlapping the
three body isospin wave function with the fact that we consider
the isoscalar channel for particles $2$ and $3$.

The same Faddeev decomposition and equations apply
for the $J^P=2^{-}$ molecule after making the substitution
\begin{eqnarray}
  && | D^{*} D^{*} \bar{D}^{*} \rangle \to \nonumber \\
  && \quad \frac{1}{\sqrt{3}}\left[
    | D^{*} D^{*} \bar{D} \rangle +
    | D^{*} D \bar{D}^{*} \rangle +
    | D D^{*} \bar{D}^{*} \rangle
    \right] \, .
\end{eqnarray}

For the $J^P=1^{-}$ $| D D \bar{D}^{*} \rangle$ molecule the inclusion of isospin
follows the same pattern as in the $J^P=3^{-}$  $| D^* D^* \bar{D}^{*} \rangle$
case. That is, we end up with the same equations as in the neutral
charmed meson case 
\begin{eqnarray}
  a_1(p_1) &=& \frac{3}{8}\tau^{\rm IS(+)}_1(Z_{23})\,
  \int \frac{d^3 p_2}{(2\pi)^3}\,
  B^0_{12}(\vec{p}_1, \vec{p}_2)\,a_1(p_2) \, , \nonumber \\ \\
  a_2(p_1) &=& \frac{3}{8}\tau^{\rm IS(+)}_1(Z_{23})\,
  \int \frac{d^3 p_2}{(2\pi)^3}\,
  B^0_{12}(\vec{p}_1, \vec{p}_2)\,a_1(p_2) \, , \nonumber \\
\end{eqnarray}
modulo the $\frac{3}{4}$ factor from the isospin projection and
the fact that we are using the isoscalar T-matrix.

\section{The Efimov Effect in the $HH\bar{H}$ System}
\label{sec:Efimov}

Now we will consider the previous set of Faddeev equations in the unitary limit.
In all cases the eigenvalue equation reduces to 
\begin{eqnarray}
  a(p_1) = \lambda \, \tau(Z_{23}) \int \frac{d^3 p_2}{(2\pi)^3}\,
  B^0_{12}(\vec{p}_1, \vec{p}_2)\, a(p_2) \, ,
  \label{eq:3B-isospin}
\end{eqnarray}
with $\lambda = 1, \frac{3}{4}, \frac{1}{2}$ and $\frac{3}{8}$
depending on the situation.
If we make the simplification $m_D = m_D^* = m_H$, which is compatible with
the heavy quark limit, and we consider the unitary limit then we have
\begin{eqnarray}
  \tau(Z_{23}) &\to& - \frac{2 \pi\,\sqrt{3}}{m_H}\,\frac{1}{p_1} \, , \\
  \int \frac{d^2 p_2}{4 \pi}\,B^0_{12} &\to& -\frac{m_H}{2 p_1 p_2}\,
  \log{\left[ \frac{p_1^2 + p_2^2 + p_1 p_2}{p_1^2 + p_2^2 - p_1 p_2} \right]}
  \, .
\end{eqnarray}
From this we arrive to the equation
\begin{eqnarray}
  p_1\,a(p_1) &=& \frac{\lambda}{\pi}\,\sqrt{\frac{3}{4}}\,\frac{1}{p_1}\,
  \int_0^{\infty} dp_2\,p_2 a(p_2)\, \nonumber \\
  && \quad \times
  \log{\left[ \frac{p_1^2 + p_2^2 + p_1 p_2}{p_1^2 + p_2^2 - p_1 p_2} \right]}
  \, ,
\end{eqnarray}
which after the change of variable $p^2 a(p) = b(p)$ transforms into
\begin{eqnarray}
  b(p) &=& \frac{\lambda}{\pi}\,\sqrt{\frac{3}{4}}\,
  \int_0^{\infty} dx \,\frac{b(x p)}{x}\,
  \log{\left[ \frac{1 + x^2 + x}{1 + x^2 - x} \right]}
  \, .
\end{eqnarray}
This equations admits power-law solutions of the type $b(p) = p^s$,
in which case we end up with an eigenvalue equation for $s$
\begin{eqnarray}
  1 &=& \frac{\lambda}{\pi}\,\sqrt{\frac{3}{4}}\,
  \int_0^{\infty} dx \,x^{s-1}\,
  \log{\left[ \frac{1 + x^2 + x}{1 + x^2 - x} \right]}
  \, ,
\end{eqnarray}
where the integral above can be evaluated analytically.
From this we can rewrite the eigenvalue equation into the more familiar form
\begin{eqnarray}
  1 = \lambda\,I_{\rm Efimov}(s) =
  \lambda\,\frac{4}{\sqrt{3}\,s}\,
  \frac{\sin{\frac{\pi s}{6}}}{\cos{\frac{\pi s}{2}}} \, .
\end{eqnarray}
The eigenvalue equation has complex solutions of
the type $s = \pm i s_0$ for
\begin{eqnarray}
  \lambda \geq \lambda_c = \frac{3 \sqrt{3}}{2 \pi} \simeq 0.826993 \, .
\end{eqnarray}
From this we can see that the $J^P = 2^{-}, 3^{-}$ three body states
might display the Efimov effect if we consider
the neutral components only, i.e.
$D^{*0} D^{*0} \bar{D}^0$ and $D^{*0} D^{*0} \bar{D}^{*0}$. 
For this configurations the solution of the previous eigenvalue equation
yields $s_0 = 0.413697$, which implies that the system shows discrete
scale invariance under transformations $p \to \mu_0\, p$ with
$\mu_0 = e^{\pi/s_0} \simeq 1986.1$.
If we consider binding energies instead of momenta the scaling transformation
becomes $E_B \to \mu_0^2\, E_B$ where $\mu_0^2 = 3.9447 \cdot 10^6$.
The sheer size of this number implies that the existence of the Efimov effect
in these systems is more of a theoretical curiosity than something
that could ever be hoped to be observed.
The energy of the first excited Efimov state of a $D^{*0} D^{*0} \bar{D}^0$ or
$D^{*0} D^{*0} \bar{D}^{*0}$ system is in fact orders of magnitude smaller
than the width of the $D^{0*}$ meson.
It is nonetheless interesting in the sense that it provides an example
of a hadronic system where this type of spectrum could be realized.

If we consider the isospin symmetric limit we have
$\lambda = \frac{3}{4} < \lambda_c$.
This is interesting in the following sense: the existence of the Efimov
effect in a three body system implies the requirement of a three-body
force to properly renormalize
the Faddeev equations~\cite{Bedaque:1998kg,Bedaque:1998km}.
That is, Faddeev calculations in the isospin symmetric limit
have predictive power.
Here it is also curious to notice that even if isospin symmetry is broken at
the level of the masses of the charmed mesons, if the interactions are
isospin symmetric there will be no requirement of three-body forces
at short distances.

For the $J^P = 1^{-}$ state we have $\lambda = \frac{1}{2}$ or $\frac{3}{8}$
depending on whether we are considering the long range neutral component
or the isospin symmetric limit.
In both cases this is insufficient to trigger the Efimov geometric scaling.
This situation is indeed equivalent to the one considered
in Ref.~\cite{Canham:2009zq} for $D^0 X$ scattering.
Here a few comments are in order: if we consider the $D^{0*} X$ system
but do not take into account that $2^{++}$ $D^{*0} \bar{D}^{*0}$
might be resonant too, then the numerical factor
for the $J^P=2^{-}$ $D^{*0} D^{*0} \bar{D}^{*0}$ system
is also $\lambda = \frac{1}{2}$ and
we end up agreeing with the conclusions of Ref.~\cite{Canham:2009zq}.
Other interesting observation is that if in the $1^{+-}$ $D^0 \bar{D}^{*0}$
system was also resonant, then we will end up with $\lambda = 1$
for the $J^P=1^{-}$ state.
Yet there is no evidence of the existence of a negative C-parity partner
of the $X(3872)$.

\section{The $J^P=2^-, 3^-$ Three Body States}
\label{sec:DDD}

\begin{table*}[ttt]
  \begin{tabular}{|c|c|c|c|cc|cc|}
    \hline
    3B System & 2B System 
    & $J^P$ & $I$ &
    $B_2$ ($0.5\,{\rm GeV}$) & $B_3$ ($0.5\,{\rm GeV}$)
    & $B_2$ ($1.0\,{\rm GeV}$) & $B_3$ ($1.0\,{\rm GeV}$) \\
    \hline \hline
    $D^* D^* \bar{D}$ & $D^*\bar{D}$/$D^* \bar{D}^*$
    & $2^{-}$ & $\frac{1}{2}$
    & Input/$5^{+3}_{-3}$ & $1.9^{+0}_{-1.4}$ & Input/$6^{+8}_{-4}$ &
    $1.3^{+0}_{-1.4}$  \\
    $D^* D^* \bar{D}^*$ & $D^* \bar{D}^*$ 
    & $3^{-}$ & $\frac{1}{2}$
    & $5^{+3}_{-3}$ & $3^{+2}_{-2}$ & $6^{+8}_{-4}$ & $3^{+4}_{-3}$ \\
    \hline \hline
    \multirow{3}{*}{$B^* B^* \bar{B}^*$} &
    \multirow{3}{*}{$B^* \bar{B}^*$} 
    & $0^-$ & $\frac{1}{2}$
    & \multirow{3}{*}{$2 \pm 2$ (Input)} &
    \multirow{3}{*}{$1.1^{+1.3}_{-1.2}$} &
    \multirow{3}{*}{$2 \pm 2$ (Input)} &
    \multirow{3}{*}{$1.0^{+1.1}_{-1.0}$} \\
    &  & $1^-$ & $\frac{3}{2}$ & & & & \\
    & & $2^-$ & $\frac{3}{2}$ & & & & \\
    \hline
  \end{tabular}
  \caption{
    Predictions for the three body binding energies of the $cc \bar c$- and
    $bb \bar b$-type molecular trimers considered in this work.
    In the upper part of the table we have the $J^P = 2^-$ $D^* D^* \bar{D}$
    and $3^-$ $D^* D^* \bar{D}$ trimers, which have been deduced
    from HQSS and the assumption that the $X(3872)$ is molecular.
    These are the most solid predictions in this work.
    In the bottom part we have the isodoublet $0^-$ and isoquartet
    $1^-$, $2^-$ $B^* B^* \bar{B}^*$ trimers.
    The binding energies of these trimers depend on the hypothesis that the
    $Z_b(10650)$ is indeed molecular and located below the $B^* \bar{B}^+$
    threshold.
    The calculations have been made in the isospin symmetric limit.
  }
  \label{tab:DDD}
\end{table*}

Here we will consider the $2^-$ and $3^{-}$ three body molecules
in the isospin symmetric limit.
In this limit we simply take the masses of the $D$ and $D^*$ charmed mesons
to be their isospin average $m_D = (m_{D^0} + m_{D^{-}}) / 2 = 1867\,{\rm MeV}$
and $m_{D^*} = (m_{D^{*0}} + m_{D^{*-}}) / 2 = 2009\,{\rm MeV}$.
As a consequence the $X(3872)$ is bound by about $4\,{\rm MeV}$.
We describe the $X(3872)$ potential in terms of a contact interaction of
the type
\begin{eqnarray}
  \langle p' | V | p \rangle = C_0(\Lambda)\,g_{\Lambda}(p') g_{\Lambda}(p) \, ,
\end{eqnarray}
with $C_0(\Lambda)$ a coupling constant that depends on a cut-off $\Lambda$
and $g_{\Lambda}$ a regulator function.
Following Ref.~\cite{Nieves:2012tt} we will take $\Lambda = 0.5-1.0\,{\rm GeV}$
and a Gaussian regulator $g_{\Lambda}(p) = \exp{(-p^2/\Lambda^2)}$.
If we now consider HQSS
in the line of Refs.~\cite{Valderrama:2012jv,Nieves:2012tt},
the $X(3872)$ implies a $2^{++}$ $X(4012)$ state with a binding energy of
\begin{eqnarray}
  B_2 = 5_{-3}^{+3}\,{\rm MeV} \quad ( B_2 = 6_{-5}^{+8} \,{\rm MeV} ) \, , 
\end{eqnarray}
for $\Lambda = 0.5\,{\rm GeV}$ ($\Lambda = 1\,{\rm GeV}$).
If we take into account isospin violation for the thresholds of the $2^{++}$
state, $D^{*0} \bar{D}^{*0}$ and $D^{*+} {D}^{*-}$, then in a first
approximation the binding energy with respect to
the neutral component is $B_2(D^{*0} \bar{D}^{*0}) \simeq B_2 - 3.3\,{\rm MeV}$.
From this we can write the condition
\begin{eqnarray}
  B_2 \geq 3.3 {\rm MeV} \, ,
\end{eqnarray}
for the neutral component to be bound and the $2^{++}$ state to be safe.
Taking into account the uncertainty in the binding energy of the $2^{++}$ state,
there is a sizable probability that its neutral component will be unbound.

For the discussion of the three body states,
we will begin with the $3^-$ $D^* D^* \bar{D}^*$ molecule.
A three body calculation indicates that the binding of the $3^-$ state is
\begin{eqnarray}
  B_3 = 3_{-2}^{+2}\,{\rm MeV} \quad (B_3 = 3_{-3}^{+4}\,{\rm MeV}) \, ,
\end{eqnarray}
below the two-body binding threshold, by which we mean that
the location of the $D^* D^* \bar{D}^*$ is $m(3^-) = 3 m_{D^*} - B_2 - B_3$.
In this case safety with respect to the isospin breaking thresholds
implies the condition
\begin{eqnarray}
  (B_2 + B_3) \geq 3.3\,{\rm MeV} \, ,
\end{eqnarray}
for the state with $| I M_I \rangle = | \frac{1}{2} -\frac{1}{2} \rangle$,
which contains the $D^{*0} D^{*0} \bar{D}^{*0}$ and $D^{*0} D^{*-} {D}^{*+}$
thresholds.
For comparison we have $(B_2 + B_3) = 8^{+4}_{-4}$ and
$9^{+9}_{-5}\,{\rm MeV}$ for $\Lambda = 0.5$ and $1\,{\rm GeV}$
(adding the errors in quadrature).
This is less stringent than in the two-body case and suggest
that there is the possibility of a Borromean configuration,
in which the two-body $2^{++}$ state is unbound
but the three body $3^{-}$ state is bound.
Yet checking these conclusions will require a full calculation including
isospin breaking, which is beyond the exploratory scope of
the present manuscript.

\begin{figure}[ttt]
  \begin{center}
\includegraphics[width=8.5cm]{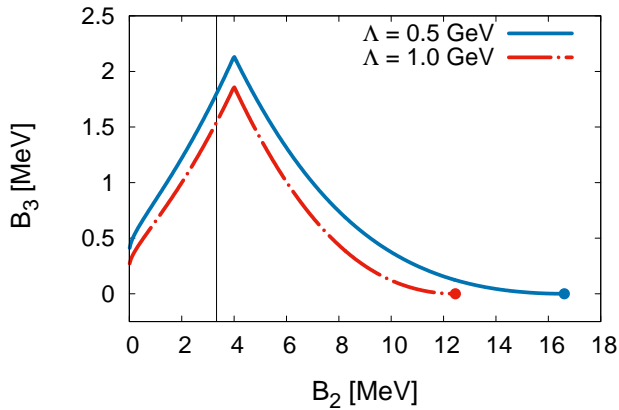}
\end{center}
\caption{
  Binding energy $B_3$ of the $J=2^-$ $D^* D^* \bar{D}$ trimer versus
  the binding energy of the theorized $J^{PC} = 2^{++}$ $D^* \bar{D}^*$
  $X_2$ partner of the $X(3872)$.
  The dots indicate the dimer binding energy for which the trimer is above
  the $X_2 D$ threshold and becomes unstable,
  which is $B_2 = 16.6$ and $12.5\,{\rm MeV}$
  for $\Lambda = 0.5$ and $1\,{\rm GeV}$ respectively.
  The vertical line indicates $B_2 = 3.3\,{\rm MeV}$:
  for $B_2 \leq 3.3\,{\rm MeV}$ the neutral $D^{*0} \bar{D}^{*0}$
  component of the $X_2$ wave function will not bind.
}~\label{fig:B2-B3}
\end{figure}

The $2^-$ $D^* D^* \bar{D}$ is more complex for the following reasons:
while the interaction in the $1^{++}$ channel is fixed,
the interaction in the $2^{++}$ channel is derived
from HQSS and subjected to a $\Lambda_{QCD} / m_c \sim 15\%$ uncertainty.
Besides, there is the technical complication that the masses of the three
charmed mesons are not identical.
This mismatch between the $1^{++}$ and $2^{++}$ channels
has a curious consequence: if the interaction in the $2^{++}$ channel
is stronger than expected, it happens that the two-body binding energy of
the $2^{++}$ $D^* \bar{D}^*$ system will grow quicker
than the three-body binding of the $2^-$ $D^* D^* \bar{D}$ system.
The reason is that the attraction increases only in one of the channels,
but not in the other.
In particular it can happen that the $2^-$ $D^* D^* \bar{D}$ bound state can
end up above the $D X_2$ threshold if the $X_2$ is too deep, which happens
for $B_2(X_2) \geq 16.6$ and $12.5 \,{\rm MeV}$
for $\Lambda = 0.5$ and $1\,{\rm GeV}$, respectively.
How this happens can be seen in Fig.~\ref{fig:B2-B3}, where the binding energy
of the $2^-$ trimer is shown as a function of the binding energy of the $X_2$.
In Fig.~\ref{fig:B2-B3} we can also notice that the most tightly bound
configuration for the trimer happens when the binding of the $X_2$
is identical to that of the $X(3872)$.
That is not the case in the $3^-$ $D^* D^* \bar{D}^*$ molecule because
the binding in the three body system grows a bit faster
than in the two body one.

The predictions for the $D^* D^* \bar{D}$ and $D^* D^* \bar{D}^*$ trimers
are summarized in Table \ref{tab:DDD}.
We stress the theoretical nature of the present work.
The $2^{++}$ partner of the $X(3872)$ has not been observed yet.
This could mean that it has simply not been detected or
it could mean that it does not exist.
Probably the best chance for its detection is
$e^+ e^{-} \to \psi(nS) \to \gamma X_2$ (with $\psi$ a $1^{--}$ charmonium)
in the $4.4-4.5\,{\rm GeV}$ region~\cite{Guo:2014ura}.
Among the theoretical reasons for the $X_2$ not to exist
the most prosaic is that HQSS is not exact: if the interaction
is weaker than expected the $X_2$ will simply
become a virtual state~\cite{Nieves:2012tt},
which might be difficult to detect.
Other possibility is that part of the attraction that binds the $X(3872)$
might come from the coupling to a nearby $1^{++}$ charmonium: the $X_2$ does
not benefit from these extra attraction, resulting in an interaction
too weak to bind~\cite{Cincioglu:2016fkm}.
In these two scenarios the three body bound states might still survive
contingent on how much attraction is lost owing to these effects.
A third option is the one pion exchange potential~\cite{Baru:2016iwj},
which could lead to a much more bound and broad $X_2$~\footnote{
This in turn might suggest the identification of the $X_2$ with
the $X(3915)$ hidden charm resonance
discovered by Belle~\cite{Uehara:2009tx}.
This was proposed for instance in Ref~\cite{Molina:2009ct},
where a detailed discussion of this scenario
can be found in Ref.~\cite{Baru:2017fgv}. A more standard
interpretation of the $X(3915)$ is that of the $\chi_{c2}(2P)$
charmonium~\cite{Zhou:2015uva}.}.
In this third scenario the $3^-$ $D^* D^* \bar{D}^*$ molecule will be also
much more bound, while the $2^{-}$ $D^* D^* \bar{D}$ trimer will decay
into $D X_2$.
Yet we consider this scenario less realistic because of the large
cut-offs required for the $X_2$ to be tightly bound~\cite{Baru:2016iwj},
while other theoretical studies with pion exchanges and a softer
cut-off find the $X_2$ to be much more shallow~\cite{Nieves:2012tt}
and narrow~\cite{Albaladejo:2015dsa}.
Be it as it may, unless the $X_2$ is detected
the discussion of the previous effects
will remain theoretical.

\section{The $J^P=0^{-}, 1^-, 2^-$ $B^* B^* \bar{B}^*$ Three Body States}
\label{sec:BBB}

The $Z_b(10610)$ and $Z_b(10650)$ isovector hidden-bottom resonances,
which we will also call $Z_b$ and $Z_b'$ for short,
are also strong candidates to be molecular.
If this is the case, they might be $B^* \bar{B}$ and $B^* \bar{B}^*$ bound
states with quantum numbers $I^G = 1^{+}$ and $J^{PC} = 1^{+-}$.
From this assumption it is easy to adapt the previous formalism
to the $B^* B^* \bar{B}^*$ system.
If we assume a non-interacting $B^* B^*$ pair, the ansatz
to the Faddeev decomposition of the wave function is
\begin{eqnarray}
  \Psi_{3B} &=&
  \left[ \phi(\vec{k}_{23},\vec{p}_1) + \phi(\vec{k}_{31},\vec{p}_2) \right] \,
  | B^* B^* \bar{B}^* \rangle \, ,
\end{eqnarray}
where the $| B^* B^* \bar{B}^* \rangle$ will be in a given spin and
isospin configuration that we have not indicated yet.
If we consider the $Z_b'$ channel to be the only non-trivial scattering
channel, the eigenvalue equation will be given by Eq.(\ref{eq:3B-isospin}).
In turn this equation depends on the numerical factor $\lambda$,
which we can write as
\begin{eqnarray}
  \lambda(B^*B^*\bar{B}^*) = \lambda_S \, \lambda_I \, ,
\end{eqnarray}
with $\lambda_S$ and $\lambda_I$ a spin and isospin factor.
The largest this factor can be is $\lambda = \frac{3}{4}$, for which
there exist three configurations of the $B^* B^* \bar{B}^*$ system.
The first is the isodoublet $J^P = 0^{-}$ configuration,
with the spin and isospin wave functions
\begin{eqnarray}
  | B^* B^* \bar{B}^* (J^P=0^-, I=\frac{1}{2}) \rangle =
  | 1_{12} \otimes 1 \rangle_J \, | 0_{12} \otimes \frac{1}{2} \rangle_I \, ,
  \nonumber \\
\end{eqnarray}
which lead to $\lambda_S = 1$ and $\lambda_I = \frac{3}{4}$.
The second and third are the isoquartet $J^P = 1^{-}$ and $2^{-}$ configurations
\begin{eqnarray}
  | B^* B^* \bar{B}^* (J^P = 1^-, I=\frac{3}{2}) \rangle &=&
  \Big( - \frac{2}{3}\,| 0_{12} \otimes 1 \rangle_J \nonumber \\ &&
  + \frac{\sqrt{5}}{3}\,| 2_{12} \otimes 1 \rangle_J  \Big)
  \nonumber \\ &\times&  | 1_{12} \otimes \frac{1}{2} \rangle_I \, ,\\
  | B^* B^* D^* (J^P = 2^-, I=\frac{3}{2}) \rangle &=&
  | 2_{12} \otimes 1 \rangle_J \, | 1_{12} \otimes \frac{1}{2} \rangle_I \, ,
  \nonumber \\
\end{eqnarray}
for which $\lambda_S = \frac{3}{4}$, $\lambda_I = 1$.
For these three configurations the eigenvalue equations are identical,
as far as we are only considering the interaction
in the $Z_b'$ channel.
The predictions depend however on what is the binding energy of
the isovector $1^{+-}$ $B^* \bar{B}^*$ molecule.
In Ref.~\cite{Cleven:2011gp} the binding energy was estimated to be
$B_2 = 4.7^{+2.3}_{-2.2}\,{\rm MeV}$ and $0.11^{+0.14}_{-0.06}\,{\rm MeV}$
for the $Z_b$ and $Z_b'$ states respectively.
Here we will simply assume
\begin{eqnarray}
  B_2 = 2 \pm 2\,{\rm MeV} \, ,
\end{eqnarray}
(the average of the previous two values) for both the $Z_b$ and $Z_b'$,
from which we deduce a three body binding energy of
\begin{eqnarray}
  B_3 = 1.1^{+1.3}_{-1.1} \, {\rm MeV} \quad (1.0^{+1.1}_{-1.0} \, {\rm MeV}) \, ,
\end{eqnarray}
for $\Lambda = 0.5\,{\rm GeV}$ ($\Lambda = 1\,{\rm GeV}$).

\begin{figure}[ttt]
  \begin{center}
\includegraphics[width=8.5cm]{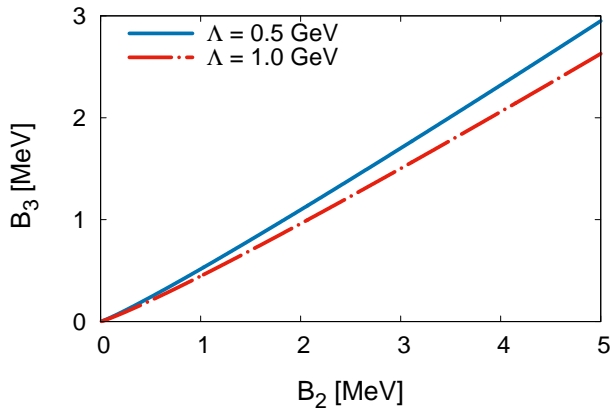}
\end{center}
\caption{
  Binding energy $B_3$ of the isodoublet $J=0^-$ $B^* B^* \bar{B}$ and
  the isoquartet $J=1^-, 2^-$ $B^* B^* \bar{B}$ trimers versus
  the binding energy of the $Z_b(10650)$ in the molecular picture,
  where it is an isovector $1^{+-}$ $B^* \bar{B}^*$ bound state,
}~\label{fig:B2-B3-Zb}
\end{figure}

The dependence of the trimer $B_3$ binding energy in the dimer $B_2$
binding energy is shown in Fig.~\ref{fig:B2-B3-Zb}.
The dependence is not linear, but can be roughly approximated
by $B_3 \sim (0.45-0.55)\,B_2$ for $B_2 \geq 1\,{\rm MeV}$.
The system is not Borromean:
if $B_2 = 0$ there will be no three body bound states.
For this reason the existence of the isodoublet $0^{-}$ $B^* B^* \bar{B}^*$
and isoquartet $1^{-}$, $2^-$ $B^* B^* \bar{B}^*$ molecules is contingent
to the exact nature of the $Z_b'$.
If the $Z_b'$ is a virtual state or if its binding is too close to
the unitary limit, it will not necessarily bind.
This happens for instance in Ref.~\cite{Wilbring:2017fwy},
where they consider $B^* Z_b'$ scattering in the dibaryon formalism:
this work finds an extremely large scattering length for the states we are
considering here, but not a bound state, where the reason lies in their
choice of the $Z_b'$ binding energy, $0.11^{+0.14}_{-0.06}\,{\rm MeV}$,
consistent with the extraction of Ref.~\cite{Cleven:2011gp}.
If we consider the most recent analysis of Ref.~\cite{Wang:2018jlv},
the $Z_b'$ would be either on top of the $B^* \bar{B^*}$
threshold or slightly above.
In the theoretical model of Ref.~\cite{Dias:2014pva} the $Z_b$'s pole is
also above the threshold.
From this the previous three $B^* B^* \bar{B}^*$ trimers should not bind,
unless there is some missing attraction not accounted for
from the channels we have not considered.
If we ignore the $B^* B^*$ interaction, this missing attraction might come
from $B^* \bar{B}^*$ scattering in spin and isospin channels different
from the $Z_b'$.
Their effect can be accounted for by changing the prefactor of
the eigenvalue equation, Eq.(\ref{eq:3B-isospin}), as follows
\begin{eqnarray}
  \lambda \tau(Z_{23}) \to \sum_{\alpha} \lambda_{\alpha} \tau_{\alpha}(Z_{23}) \, ,
\end{eqnarray}
where $\alpha$ refers to the other possible
$B^* \bar{B}^*$ scattering channels.
The missing channels for each of the trimers are:
\begin{enumerate}
\item[(i)] For the isodoublet $J^P = 0^{-}$ trimer we have
  the $I^G(J^{PC})$ = $0^{-}(1^{+-})$ channel with $\lambda = 1/4$.
\item[(ii)] For the isoquartet $J^P = 1^{-}$ trimer we have
  the $I^G(J^{PC})$ = $1^{-}(0^{++})$ and $1^{-}(2^{++})$ channels
  with the factors $\lambda = 1/9$ and $5/36$, respectively.
  \item[(iii)] For the isoquartet $J^P = 2^{-}$ trimer we have
  the $I^G(J^{PC})$ = $1^{-}(2^{++})$ channel with $\lambda = 1/4$.
\end{enumerate}
We will merely mention the existence of these channels,
but will not take them into account.
We do not have any experimental information about them and their treatment
will require a phenomenological model of their interactions.
However we will mention that bound states have been theorized
in a few of these channels~\cite{Voloshin:2011qa,Mehen:2011yh,Baru:2017gwo}.

\section{The $J^P=0^{-}$ $BBD$, $1^-$ $BBD^*$ and $2^-$ $B^*B^* D$
  Three Body System}
\label{sec:BBD}

\begin{table*}[ttt]
  \begin{tabular}{|c|c|c|c|cc|cc|}
    \hline
    3B System & 2B System 
    & $J^P$ & $I$ &
    $B_2$ ($0.5\,{\rm GeV}$) & $B_3$ ($0.5\,{\rm GeV}$)
    & $B_2$ ($1.0\,{\rm GeV}$) & $B_3$ ($1.0\,{\rm GeV}$) \\
    \hline \hline
    $BBD$ & $BD$ 
    & $0^{-}$ & $\frac{1}{2}$
    & \multirow{3}{*}{$4^{+3}_{-2}$} 
    & \multirow{3}{*}{$4^{+2}_{-2}$}
    & \multirow{3}{*}{$5^{+6}_{-4}$} 
    & \multirow{3}{*}{$6^{+5}_{-3}$} \\
    $BB^*D$ & $BD$/$B^*D$ 
    & $1^{-}$ & $\frac{1}{2}$
    & & & & \\
    $B^* B^* D$ & $B^*D$ 
    & $2^{-}$ & $\frac{1}{2}$
    & & & & \\
    $B B D^*$ & $BD^*$ 
    & $1^{-}$ & $\frac{1}{2}$
    & $5^{+3}_{-3}$ & $4^{+2}_{-2}$ & $7^{+7}_{-5}$ & $7^{+6}_{-5}$ \\
    \hline \hline
    \multirow{3}{*}{$B^* B^* D^*$} &
    \multirow{3}{*}{$B^* D^*$} 
    & $0^-$ & $\frac{1}{2}$
    & \multirow{3}{*}{$\dagger_{-0.6}$} & \multirow{3}{*}{$0.2^{+\dagger}_{-0.6}$}
    & \multirow{3}{*}{-} & \multirow{3}{*}{-}  \\
    & & $1^-$ & $\frac{3}{2}$ & & & & \\
    & & $2^-$ & $\frac{3}{2}$ & & & & \\
    \hline
  \end{tabular}
  \caption{
    Predictions for the three body binding energies in the isospin symmetric
    limit for the mass-imbalanced $\bar{b} \bar{b} c$-type molecular trimers.
    In the upper part of the table we have the isodoublet
    $J^P = 0^-$ $B B D$, $1^-$ $B B^* D$, $2^-$ $B^* B^* D$ and
    $1^-$ $B B D^*$ molecules.
    Their binding is derived from the assumption that the $X(3872)$
    is molecular, heavy quark symmetry (HQSS and HFS)
    and the hypothesis that the two-body $BD$ interaction
    in the light-spin $S_L = 0$ channel
    is negligible in comparison with the $S_L = 1$ one.
    In the bottom part there are the predictions for the $B^* B^* {D}^*$
    partners of the $B^* B^* \bar{B}^*$ trimers,
    for which we have used HFS.
    The binding of these trimers is however dependent on the cut-off
    used in the calculations.
  }
  \label{tab:BBD-trimers}
\end{table*}

The heavy-light spin decomposition of the $1^{++}$ $D\bar{D}^*$
and the $2^{++}$ $D^* \bar{D}^*$ states is  $1_H \otimes 1_L$,
as previously explained.
This leads to a potential of the type
\begin{eqnarray}
  V(D\bar{D}^*, 1^{++}) = V(D^*\bar{D}^*, 2^{++}) = V_1 \, ,
\end{eqnarray}
where $V_1$ indicates the potential for $S_L = 1$,
with $S_L$ the total light spin.
The particular decomposition depends on the symmetries and quantum numbers
of the heavy-light system under consideration.
If we consider the $BD$, $B^*D$, $BD^*$ systems the heavy-light decomposition
implies a potential of the type~\cite{Mehen:2011yh,Valderrama:2012jv}
\begin{eqnarray}
  V(BD) = V(B^* D) = V(D^* B) &=& \frac{1}{4}\,V_0 + \frac{3}{4}\,V_1 \, ,
\end{eqnarray}
where now there is $V_0$, the potential for $S_L = 0$.
For the $B^* D^*$ system the decomposition will depend
on the $J^P$ quantum numbers
\begin{eqnarray}
  V(B^*D^*, J=0^+) &=&  \frac{3}{4}\,V_0 + \frac{1}{4}\,V_1 \, , \\
  V(B^*D^*, J=1^+) &=&  \frac{1}{2}\,V_0 + \frac{1}{2}\,V_1 \, , \\
  V(B^*D^*, J=2^+) &=&  V_1 \, .
\end{eqnarray}
According to heavy flavour symmetry (HFS), the potentials $V_0$ and $V_1$
are identical for the $c\bar{c}$ and the $c\bar{b}$ sectors~\cite{Guo:2013sya}.
Hence we can relate the $B D$/$B^* D$/$\dots$ potentials
with the $\bar{D} D$/$\bar{D}^* D$/$\dots$ ones.

From the $c\bar{c}$ sector we know that $V_1$ is strong,
but we do not know that much about $V_0$.
However in the isoscalar sector there is a conspicuous experimental absence of
other hidden charm molecular $D\bar{D}$/$D^*\bar{D}$/$D^*\bar{D}^*$
candidates besides the $X(3872)$.
This points out to a $V_0$ that is either repulsive or weaker than $V_1$.
From this we will make the assumption that $|V_0| \ll |V_1|$
and explore the consequences.
We warn that this assumption is based on very partial information.
As a matter of fact there are theoretical models in which
the $V_0$ interaction is indeed as attractive as, if not more attractive than,
the $V_1$, leading to the prediction of multiplets of
six hidden charm molecular states~\cite{Gamermann:2006nm,Nieves:2012tt,HidalgoDuque:2012pq}.
But here we will simply explore the consequences of $V_0 = 0$.

Independently of $V_0$, there is the trivial conclusion that
the $2^+$ $B^* D^*$ molecule should bind~\cite{Guo:2013sya}.
The rationale is that attraction in non-relativistic bound states depends on
the reduced potential $U = 2 \mu V$, with $\mu$ the reduced mass
and $V$ the potential.
For the $2^+$ $B^* D^*$ molecule the potential is identical to the one for
the $1^{++}$ $D\bar{D}$ molecule, i.e. the $X(3872)$,
and the reduced mass is $1.51$ times larger.
Binding is expected and concrete calculations indicate
a bound state at $B = 12 \pm 5\,{\rm MeV}$ ($26^{+14}_{-13}\,{\rm MeV}$)
for $\Lambda = 0.5\,{\rm GeV}$ ($1.0\,{\rm GeV}$)~\cite{Guo:2013sya}.
The $BD$, $B^* D$ and $B D^*$ cases are more interesting.
If $V_0 = 0$ the strength of the $BD$, $B^* D$ and $B D^*$ potentials
is $3/4$ of the one in the $X(3872)$, while their reduced masses is
a bit above $4/3$ of the one in the $D^*\bar{D}$ system.
Concrete numbers show
\begin{eqnarray}
  \frac{3}{4}\,\frac{2 \mu_{DB}}{2 \mu_X}  \simeq 1.07 \, ,
\end{eqnarray}
indicating that the non-relativistic description of the two systems
should be similar.
In particular we expect the $D^0 B^+$, $D^0 B^{*+}$ and $D^{*0} B^+$
scattering lengths to be unnaturally large.

This opens the possibility of having the Efimov effect in the three body case.
If we consider the $BBD$, $B^*BD$/$BB^*D$, $BBD^*$ and $B^*B^*D$ systems,
the Faddeev equations are analogous to the ones for
the $D^* D^* \bar{D}$ and $D^* D^* \bar{D}^*$ cases. 
We begin with the ansatz
\begin{eqnarray}
    \Psi_{3B} = \left[ \phi(\vec{k}_{23},\vec{p}_1) + \phi(\vec{k}_{31},\vec{p}_2)
      \right] \, | H_{\bar b} H_{\bar b} H_{c} \rangle \, ,
\end{eqnarray}
where the heavy meson wave function refers to one of these possibilities
\begin{eqnarray}
  | H_{\bar b} H_{\bar b} H_{c} \rangle &=& | B^{+} B^{+} D^0 \rangle \, , \\
  | H_{\bar b} H_{\bar b} H_{c} \rangle &=&
  \frac{1}{\sqrt{2}}| B^{*+} B^{+} D^0 \rangle +
  \frac{1}{\sqrt{2}}| B^{+} B^{*+} D^0 \rangle
  \, , \nonumber \\ \\
  | H_{\bar b} H_{\bar b} H_{c} \rangle &=& | B^{*+} B^{*+} D^0 \rangle \, , \\
  | H_{\bar b} H_{\bar b} H_{c} \rangle &=& | B^{+} B^{+} D^{*0} \rangle \, ,
\end{eqnarray}
depending on the molecule considered.
In either case, following the same steps as before,
we end up with the equation
\begin{eqnarray}
  a(p_1) = \tau(Z_{23})\,\int \frac{d^3\vec{p}_1}{(2\pi)^3}\,
  B_{12}(\vec{p}_1,\vec{p}_2)\,a(p_2) \, .
\end{eqnarray}
The difference is that there is now a mass imbalance.
If we consider the unitary limit, particularizing for the present case,
we now obtain the equation
\begin{eqnarray}
  1 = \lambda\,J_{\rm Efimov}(s, \alpha) \, ,
\end{eqnarray}
where $J_{\rm Efimov}$ is known~\cite{Helfrich:2010yr,Helfrich:2011ut}
\begin{eqnarray}
  J_{\rm Efimov}(s, \alpha) = \frac{1}{\sin{2 \alpha}}\,\frac{2}{s}\,
  \frac{\sin{\alpha s}}{\cos{\frac{\pi}{2} s}} \, ,
\end{eqnarray}
and with the angle $\alpha$ determined as
\begin{eqnarray}
  \alpha = \arcsin{\left( \frac{1}{1 + \frac{m_3}{M}} \right)} \, ,
\end{eqnarray}
with $m_3$ the mass of the charmed meson and $M$ the mass of the bottom meson.
We find that the condition for having the Efimov geometric spectrum is
\begin{eqnarray}
  \lambda \geq \lambda_c = \frac{\sin{2 \alpha}}{2 \alpha} \, , 
\end{eqnarray}
which for the $BBD$ case give us $\lambda_c \simeq 0.599$ plus similar
values for the other cases under consideration.
It is also interesting to notice that for $m_3 = M$ we have $\alpha = \pi/6$
and we recover the original Efimov integral
\begin{eqnarray}
  I_{\rm Efimov}(s) = J_{\rm Efimov} (s, \frac{\pi}{6}) \, .
\end{eqnarray}

\begin{table}[ttt]
  \begin{tabular}{|cccccccc|}
    \hline
    System & $J^P$ & $I$ & $M / m$ & $\lambda_c$
    & $s_0$ & $e^{\pi/s_0}$ & $e^{2\pi/s_0}$ \\
    \hline \hline
    $B^+ B^+ D^0$ & $0^-$ & - & $2.83$ & $0.599$ & $0.7513$ & $65.48$ & $4287.3$ \\
    $B^{*+} B^{*+} D^0$ & $2^-$ & - & $2.86$ & $0.597$
    & $0.7546$ & $64.27$ & $4131.3$ \\
    $B^+ B^{+} D^{*0}$ & $1^-$ & - & $2.63$ &$0.614$
    & $0.7263$ & $75.61$ & $5716.8$ \\
    \hline \hline
    $B B D$ & $0^-$ & $\frac{1}{2}$ & $2.83$ & $0.599$
    & $0.4691$ & $810.0$ & $6.561 \cdot 10^5$ \\
    $B^{*} B^{*} D$ & $2^-$ & $\frac{1}{2}$ & $2.85$ & $0.597$
    & $0.4731$ & $765.2$ & $5.885 \cdot 10^5$ \\
    $B B D^{*}$ & $1^-$ & $\frac{1}{2}$ & $2.63$ &$0.614$ & $0.4345$
    & $1381.8$ & $1.909 \cdot 10^6$ \\
    \hline \hline
    $B^* B^* D^{*}$ & $0^-$ & $\frac{1}{2}$ & $2.65$ &$0.614$ & $0.4385$
    & $1291.9$ & $1.669 \cdot 10^6$ \\
    $B^* B^* D^{*}$ & $2^-$ & $\frac{3}{2}$ & $2.65$ &$0.614$ & $0.4385$
    & $1291.9$ & $1.669 \cdot 10^6$ \\
    \hline
  \end{tabular}
  \caption{
    Candidate $\bar b \bar b c$-like three heavy meson molecules
    for which the Efimov effect might be possible.
    In the top part of the table we consider systems for which
    $B^+D^0$, $B^+D^{*0}$ and $B^{*+}D^0$ scattering might be resonant
    if the interaction in the isoscalar $S_L = 0$ channel is
    considerably weaker than in the $S_L = 1$ one.
    In the middle part we consider the same systems
    in the isospin symmetric limit.
    In the bottom part we consider systems for which the Efimov effect could
    be present if the isovector $J^P = 1^+$ $B^* D^*$ scattering is resonant,
    where this channel is the $\bar b c$ heavy flavour partner of
    the $Z_b$ and $Z_b'$ hidden bottom states.
    In the table above $J^P$ is the spin and parity of the state,
    $I$ the isospin (if well-defined), $M/m$ is the mass imbalance of
    the system, $\lambda_c$ the critical coupling for the Efimov effect
    to appear, $s_0$ the power-law scaling, $e^{\pi / s_0}$ the discrete
    scaling factor for the momenta and $e^{2 \pi / s_0}$ the discrete
    scaling factor for the three-body binding energies.
  }
  \label{tab:BBD-a}
\end{table}

If we go back to the $B^+ B^+ D^0$ three body system, we have $\lambda = 1$
and there should be discrete scale invariance in the unitary limit.
If we define $\mu_0 = e^{\pi / s_0}$ we find the value $\mu_0 = 65.6$ and
$\mu_0^2 = 4310$ for the $B^+ B^+ D^0$ molecule,
plus similar numbers for the other cases,
see Table \ref{tab:BBD-a} for details.
In the isospin symmetric limit we have $\lambda = \frac{3}{4}$,
which is now strong enough as to trigger the Efimov effect,
though in this case the effect will be really weak.
However in the isospin symmetric limit we do not expect the $BD$ system
to be close enough to the unitary limit as to display the Efimov effect.
Yet it is important because this involves the existence of a three body force
at least in the $\Lambda \to \infty$ limit~\cite{Bedaque:1998kg,Bedaque:1998km}.
This however will happen at ridiculously high cut-offs (clearly beyond
the breakdown scale of the theory), which means that
we can effectively estimate the binding energy of the $0^-$ $B B D$ to be
\begin{eqnarray}
  B_3 = 4 \pm 2 \, {\rm MeV} \quad (B_3 = 6^{+5}_{-3} \, {\rm MeV}) \, ,
\end{eqnarray}
for $\Lambda = 0.5\,{\rm GeV}$ ($\Lambda = 1.0\,{\rm GeV}$).
The predictions for the $1^-$ $B B^* D$ and $2^-$ $B^* B^* D$ are identical,
as a consequence of the small mass difference between the $B$ and $B^*$ mesons.
Meanwhile the predictions for the $1^-$ $BBD^*$ trimer are slightly more bound
for the $\Lambda = 1.0\,{\rm GeV}$ case.
In Table~\ref{tab:BBD-trimers} we include a list of these states,
their quantum numbers and their properties.

If we consider the heavy meson-antimeson interaction in the isovector channel,
then it appears that the existence of the $Z_c(3900)$ and $Z_c'(4020)$
in the hidden charm sector can indeed be deduced from HFS and
the $Z_b(10610)$ and $Z_b'(10610)$
in the hidden bottom sector~\cite{Guo:2013sya}.
This argument also predicts the existence of an isovector $B^* D^*$
bound or virtual state near the threshold with $J^P = 1^{+}$.
If this prediction were to be confirmed.
it would open the possibility of Efimov physics for the $B^* B^* D^*$ system.
We end up with the same configurations, equations and isospin factors than
in the previous section: $0^-$ isodoublet and $1^-$, $2^-$ isoquartet.
The only difference is the mass imbalance
from the $\bar{B}^* \to D^*$ substitution.
The discrete scaling factor is $\mu_0 = e^{\pi/s_0} = 1291.9$ in this case,
leading to a geometric factor of $1.669 \cdot 10^6$ for the spacing of
the excited Efimov states, see Table~\ref{tab:BBD-a}.
As in the previous case, concrete calculations require a three body force.
If we ignore this requirement there is the possibility of a very shallow
trimer for $\Lambda = 0.5\,{\rm GeV}$ which disappears at larger cut-offs,
see Table~\ref{tab:BBD-trimers}.

Finally we stress that the conclusions of this section are rather theoretical.
They depend on a series of assumptions to be correct and on a series of
theoretical uncertainties to lean into the right direction.

\section{Conclusions}
\label{sec:Conclusions}

In this work we have considered
the $J^P=2^-$ $D^* D^* \bar{D}$ and $3^-$ $D^* D^* \bar{D}^*$ molecules.
From HQSS we expect the binding energy and properties of
these two systems to be identical.
Calculations in the isospin symmetric limit indicate a
three body binding energy of $B_3 \sim 1.5\,{\rm MeV}$ and $3\,{\rm MeV}$
for the $2^-$ and $3^-$ trimers, respectively.
The rationale behind this prediction is straightforward:
the application of HQSS in the charmed meson-antimeson system implies that
the potential for the $1^{++}$ $D\bar{D}^*$ and
$2^{++}$ $D^* \bar{D}^*$ systems is the same.
The $1^{++}$ $D\bar{D}^*$ can be identified with the $X(3872)$,
from which we can deduce the strength of the potential.
In the isospin-symmetric limit
$2^-$ $D^* D^* \bar{D}$ and $3^-$ $D^* D^* \bar{D}^*$ molecules can be computed
in terms of this two body potential,
provided that the $DD^*$ and $D^*D^*$ interaction is weak,
as seems to be the case from phenomenological considerations.
The $2^-$ and $3^-$ molecules do bind indeed, leading
to the previous predictions.
If instead of the $X(3872)$ and the $D^* D^* \bar{D}$ and $D^* D^* \bar{D}^*$
systems we consider the $Z_b(10650)$ as an isovector $1^{+-}$ $B^* \bar{B}^*$
molecule, then we can determine whether there are isodoublet $0^-$ and
isoquartet $1^-$ and $2^-$ $B^* B^* \bar{B}^*$ trimers.
The existence of these trimers depends on the $Z_b(10650)$ being a bound state
instead of a virtual state or resonance.

We have also investigated the conditions for the existence of the Efimov effect.
If we consider the neutral components of these systems,
i.e. the $D^{0*} D^{0*} \bar{D}^0$ and $D^{0*} D^{0*} \bar{D}^{0*}$ molecules,
and if the $D^{0*} \bar{D}^{0*}$ scattering length is unnaturally large,
then the Efimov geometric spectrum is in principle possible.
The relevance of this possibility is mostly theoretical though:
the discrete scaling factor is about $1986.1$ and the first Efimov state
should be four millions times less bound than the fundamental state.
This makes the existence of a geometric spectrum more of a theoretical nicety
than a phenomenon that we could realistically ever expect to observe,
except in the lattice perhaps.
The Efimov effect is absent in the isospin symmetric limit.

A more promising candidate for the Efimov effect are
the family of $0^-$ $B^+ B^+ D^0$, $1^-$ $B^{*+} B^+ D^0$,
$2^-$ $B^{*+} B^{*+} D^0$ and $1^-$ $B^{+} B^{+} D^{*0}$ molecules.
In these systems there is a marked mass imbalance between the bottom
and charmed mesons that favors the appearance of Efimov physics.
From HFS there is the possibility that $B^{+} D^0$ scattering might be resonant,
which in turn will imply the existence of an Efimov spectrum
for the aforementioned molecules.
The discrete scaling factor is between $65-70$, indicating that the binding of
the excited Efimov state should be $4000-5000$ times shallower than
the fundamental state.
Independently of this, it is worth noticing that three body systems
with a large mass imbalance are more likely to bind, as illustrated with
the $B^* B^* \bar{K}$~\cite{Valderrama:2018knt},
$D^*\bar{D}^*\rho$~\cite{Ren:2018pcd},
$D^*\bar{D}^*\rho$~\cite{Bayar:2015oea},
$B^*\bar{B}^*\rho$~\cite{Bayar:2015zba} and $B \bar{B} D$~\cite{Dias:2018iuy}
systems.

We stress the theoretical nature of the present work.
The $2^{++}$ partner of the $X(3872)$ has not been observed yet
neither ruled out.
Its existence has been extensively discussed
in the literature~\cite{Valderrama:2012jv,Nieves:2012tt,HidalgoDuque:2012pq,Albaladejo:2015dsa,Cincioglu:2016fkm,Baru:2016iwj}.
Without knowing whether this state exists or its binding,
it is difficult to make definite predictions about
prospective three body states.
Nonetheless the possibility of a Borromean $3^-$ $D^* D^* \bar{D}^*$ molecule
is there, which if detected will provide relevant information
about the $D^* \bar{D}^*$ interaction.
The predictions for the $\bar{b} \bar{b} b$- and $\bar{b} \bar{b} c$-type
molecules are more hypothetical.
The $\bar{b} \bar{b} b$ trimers are conditional to the location
of the $Z_b(10650)$ or the existence of HQSS partners,
while the $\bar{b} \bar{b} c$ trimers rely on the assumption
that the heavy meson-antimeson potential for the $S_L = 0$ configuration
is weaker than for the $S_L = 1$ case, where $S_L$ is the total light spin.
Yet the eventual observation of a shallow $B^+ D^0$ molecule will be really
exciting owing to its connection with Efimov physics.

\section*{Acknowledgments}

This work is partly supported by the National Natural Science Foundation
of China under Grants No.11522539, No.11735003, the Fundamental
Research Funds for the Central Universities and the Thousand
Talents Plan for Young Professionals.


\end{document}